# Stay with Your Community: Bridges between Clusters Trigger Expansion of COVID-19

Yukio Ohsawa[1*], Masaharu Tsubokura[2]

[1] Department of Systems Innovation, School of Engineering, The University of Tokyo, Tokyo, Japan

[2] Department of Public Health, Fukushima Medical University School of Medicine, Fukshima, Japan

* Corresponding author E-mail: ohsawa@sys.t.u-tokyo.ac.jp (YO)

# Abstract

The spreading of virus infection is here simulated over artificial human networks. Here, the real-space urban life of people is modeled as a modified scale-free network with constraints. A scale-free network has been adopted in several studies for modeling on-line communities so far but is modified here for the aim to represent peoples' social behaviors where the generated communities are restricted reflecting the spatiotemporal constraints in the real life. Furthermore, the networks have been extended by introducing multiple cliques in the initial step of network construction and enabling people to contact hidden (zero-degree) people as well as popular (large degree) people. As a result, four findings and a policy proposal have been obtained. First, the "second waves" have been observed in the simulations even without external influence or constraints on peoples' contacts or the releasing of the constraints. These second waves tend to be lower than the first wave and to be caused in the "fresh" clusters i.e. by the infection of people who are connected in the network but had not been infected. This implies the bridges between infected and fresh clusters may trigger new expansions of spreading. Second, if the network changes the structure on the way of infection spreading or after its suppression, the peak of the second wave can be larger than the first. Third, the peak height in the time series of the number of infection cases depends on the difference between the upper bound of the number of people each member accepts to meet and the number of people one chooses to meet. This tendency is observed for two kinds of artificial networks introduced here and implies the impact of the bridges between communities on the virus spreading. Fourth, the release of once given constraint may trigger a second wave higher than the peak of the time series without introducing any constraint from the beginning, if the release is introduced at a time close to the peak. Thus, all in all, both the government and individuals should be careful in returning to human society with inter-community contacts.

# Introduction

Predicting the spread of infection during a pandemic and taking appropriate precautions are critical components in finding the best solution to a major public health challenge. The novel coronavirus (SARS-CoV-2), which was reported in Wuhan, Hubei Province, China around December 2019, spread rapidly throughout the world, causing Covid-19 disease. The disease was relatively quickly declared a pandemic by the World Health Organization (WHO) and, as of May 2020, over 375,000 deaths had occurred and more than 6.2 million cases in 110 countries had been confirmed. In this context, various measures for the prevention of infection have been presented by the authorities in each country, mostly following WHO recommendations. In addition to personal preventive measures such as handwashing and wearing face masks, social preventive measures such as physical distancing and drastically restricted individual movements (lockdowns) were implemented. Such preventive measures were implemented by several authorities, with varying degrees of success, depending on how well they were implemented. However, the uninformed still ask questions such as "Is the disease spread suppressed as expected when people reduce their contacts with others? If so, whom should we avoid to contact?"

In this paper, we detail a model based on a simulation of infection spreading in a social network of people in an urban environment for obtaining effective prevention measures to answer such questions as above. The reason why we use the social network model instead of the population models with differential equations such as the SIR models is that we aim to clarify the effects of actions of people that generate, prune, or perturb local connections with neighbors so that we acquire knowledge about measures for preventing infection spread. This benefit of the network-based model has been pointed out in [1], where the difference from population-based models (such as SIR [2] and its extensions [3-7]), the effects of preventive actions such as a lock-down and its release and of infections via edges bridging far-apart nodes on infection spread has been evaluated.

Various network models of infection have been built to predict the spread of infection and allow informed decision making with regard to control and prevention [8–12]. For example, in [8], the dynamics have been modeled to explain the increase of infections in the early stage based on the power-law followed by exponential suppression. This model has been used and extended to obtain an indicator of a failure of the current containment efforts or the beginnings of the end of the pandemic [9], and to compare the temporal trends of daily cases in countries [10]. Strategies for network manipulation have also been explored, one of which is the control of the dynamics of disease spread via interacting with features of a disease network, such as infection rate [11], while another is network optimization by removing/rewiring links that represent an NP-hard problem [12]. The solutions for these problems include a top-down (e.g. to suppress infection rates) and a bottom approach (e.g. to enable individuals to discern the critical in-network contacts). Thus, working on network dynamics contributes to providing points to consider in designing policies for controlling epidemic spreading. The guidelines for attention and actions obtained from such research could be useful for designing policies for controlling epidemics in a timely fashion.

A lot is known about general theories applicable to various propagation phenomena in a social network. For example, it has been found that the cascades in skewed human networks are triggered by large degree nodes, whereas small degree nodes may trigger cascades in other networks [13]. In [14], the diffusion of innovation has been simulated in artificial social networks resulting in delayed propagation to low-degree nodes from high-degree nodes, which is of a greater impact if informative and nominal contacts from neighboring nodes are mixed. The authors' express an interest in the mutual influence between high-degree nodes and low-degree nodes, in which the role of the latter becomes focused. An encouraging trend nowadays is that the authors of [9] and [14], as well as others working on network dynamics, are contributing valuable information to help determine optimal intervention strategies [15].

In our simulated society, we assume that people become interested in attending a place where a lot of other individuals gather. This action may be directed not only to other people in the place in particular, but also to something that is of special attraction, either abstract or real. In this place, a new person attending may come into close contact with other humans, touch objects, and breathe the air in the place. These are similar elements seen in the social dynamics known to form a scale-free network [16] where new nodes get connected to high-degree nodes. However, there are real-space and real-time constraints on the dynamics that are 1: the restriction of time and the capacity of one's interest ($W$: the number of other people each person can meet after all), 2: the restriction of space ($W$ may be interpreted as the area or volume of a meeting place), 3: the interest in meeting other people ($m_0$: the number of other people each person meets by choice, either singly or in a group). In this paper, we aim to find some clues to further the discussion using simple simulations involving two kinds of constrained scale-free networks (SFN). From the simulated cases varying $W$ and $m_0$, we reach a finding of the effects of bridges between communities on the infection expansion that is intuitively understandable to uninformed people who desire a preventive measure an individual can take.

# Materials and methods
## The hypothesis

Each snapshot in Figure 1 shows a social network. Each node represents a human, a place, or a thing in the place. Each edge represents a possibility that the entities represented by the two nodes can interact. Each cluster in Figure 1 with thick solid edges means a group of people who choose to meet each other. People in such a group are connected mutually via strong ties, whereas the thin lines represent weaker ties between the groups via peoples' personal interests in external activities. Here, the red nodes represent people who have the current infectivity to others who are chosen randomly by probability $p$. The blue nodes have not yet infected, and the green nodes represent people who lost the infectivity to others.

Here, we set the following four hypotheses based on networked models of society.

*Hypothesis 1*: Heights of the peaks of the number of infection cases in the time sequence depend on the upper bound of the number of people each member accepts to meet ($W$), and the curve of this dependency varies on the number of people one chooses to meet ($m_0$).

*Hypothesis 2*: The new waves, e.g., the second wave, of infection spreading may occur without external events such as an increase of in-bound travelers or a release of the governmental constraints on peoples' contacts.

*Hypothesis 3*: The change in the social structure i.e., each person's choice of whom to meet, can be a trigger of a wide infection spreading.

*Hypothesis 4*: The release of the constraints on peoples' contacts can cause a wide infection spreading, if the release is introduced too early.

Hypothesis 1 can be understood by viewing the simplest Scenario A in Fig 1. As shown in the top-right of Fig.1, $W$ corresponds to all the edges connected to the member, and $m_0$ corresponds to the number of thick solid edges in a cluster (minus one as in Fig.1, for a cluster to be obtained in the initial step in the Algorithms 2 and 3 mentioned later). A cluster can be regarded as a community of people. The value of $W$–$m_0$ represents the width of the bridge between communities corresponding to the number of thin edges between clusters in Fig.1 and plays the role to enable virus infections from one to other communities. For a larger $W$– $m_0$, the peak of infection cases increases because the infection spreads via strong tie quickly.

Hypothesis 2 corresponds to Scenario B in Fig 1. Here, due to a small extension of reproductive period of the arrowed node, the virus gets infected to new communities. As a result, people (nodes) in a new community get infected, which results in the appearance of a new wave. Hypothesis 3 corresponding to Scenario C of Fig 1 considers the movement of edges to connect infected people having the infectivity and non-infected people who have not yet acquired immunity. Hypothesis 4 corresponds to Scenario D1 and D2, where the restriction on the contacts of people is released and causes edges to emerge. If time has passed after the peak of infection spread by the time of release, as in Scenario D1, the expansion will not be as severe as in Scenario D2 because a certain portion of people in the community have already been infected. In Scenario D2, the restriction is released at a time when fresh (most part is free from infection yet) clusters still exist with links to infected people.

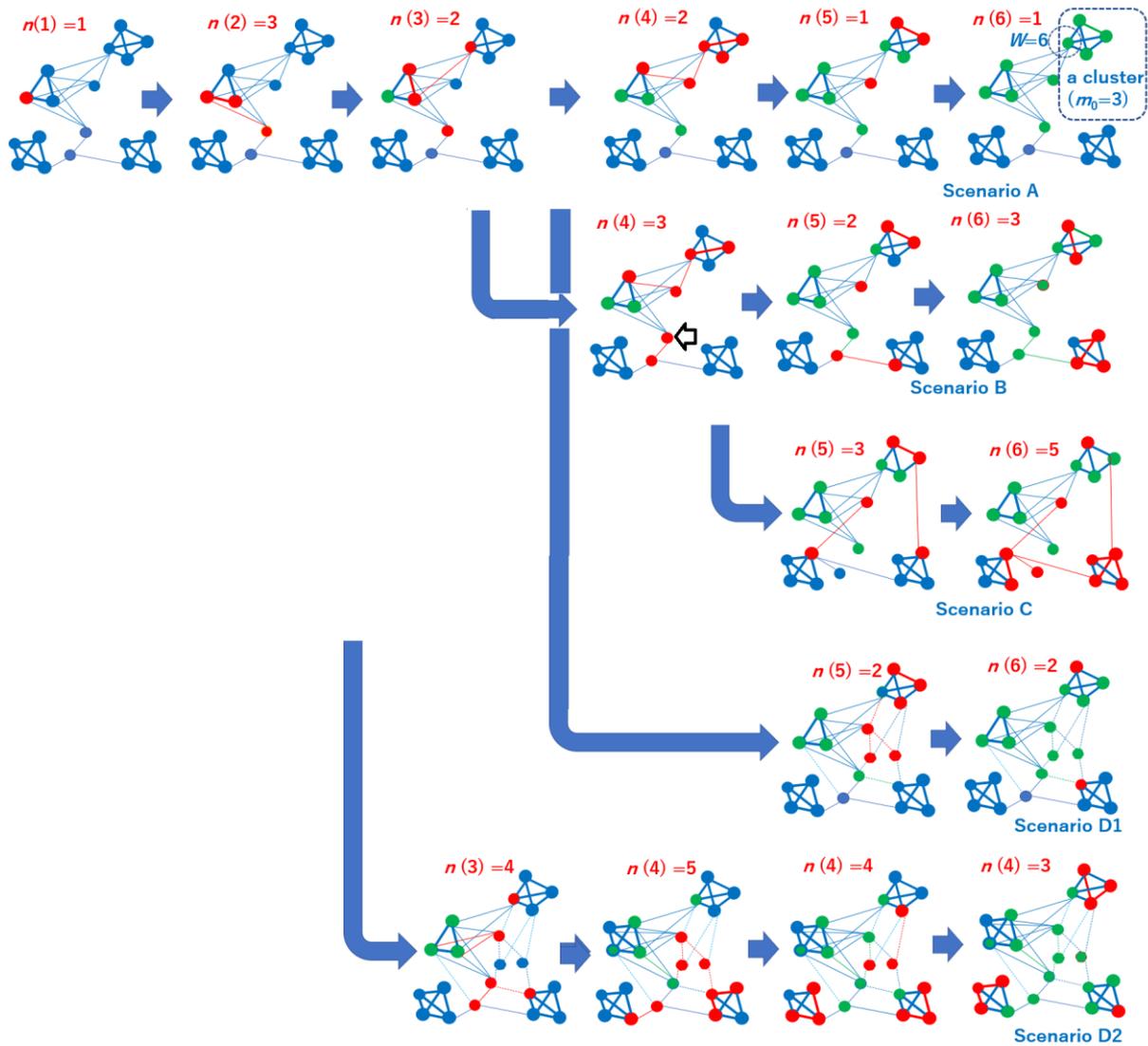

**Fig 1. The four scenarios of infection spreading in a social network. This figure shows a simplified conceptual graph. For example, the clusters generated initially should be smaller than ones generated later in Algorithms 2 and 3 mentioned later.**

## Scale-free network as the backbone

The established model to generate a scale-free network (SFN) [16] is described as follows. Hereafter, $N$ means the number of nodes to be finally included in the society. $V$ and $E$ mean the sets of nodes and edges in the graph respectively, of which the combination is defined to be the graph $G$. The graph $G$ starts from a clique of $m_0$ nodes and incremented by adding one node_$i$ at each time. Then nodes other than $m_0$ nodes are chosen as new link destinations from node_$i$ according to the probabilities given in proportion to the node degrees. Here, $m_0$ is the initial degree of each node (the number of edges connected to a new node), and the ranking in line 8 of the Algorism 1 means the rank (in $\{0, 1, 2, …\}$) of node_$j$ in choosing on the probability given in proportion to its degree denoted deg(node_$j$). That is, the values of deg(node_$j$), multiplied by a random value between 0 and 1, are compared for ranking the nodes in $G$ to choose the top $m_0$ nodes. In the real human society, $m_0$ can be regarded as the number of people whom each person intentionally chooses to meet i.e., all other people one can meet separately if preferable, not necessarily in a group. After the intended meetings, one may get contacted by other nodes each of which (who, if the node is a human) chooses $m_0$ nodes to meet.

**Algorithm 1: the generation of a scale-free network (SFN)**
1: $G := \{V, E\}$, $E = \{\}$
2: $V = \{\text{node}_1, \text{node}_2, ..., \text{node}_m\}$
3: $E \leftarrow E + \{\text{edge}_{i,j}\}$ **for all** $\text{node}_i \in V_k$, $\text{node}_j \in V_k$ ($i \neq j$)
4: **for** $i = m+1 : N$ **do**
5:     **add** $\text{node}_i$ **to** $V$
6:     $\deg(\text{node}_i) = m$
7:     **for** $j = m+1 : N$   ($i \neq j$) **do**
8:         **if** $rank_{\text{node}_j \in V} \text{random}(1) \deg(\text{node}_j) \leq m$
9:             $E \leftarrow E + \{\text{edge}_{i,j}\}$
10:             $\deg(\text{node}_j) \leftarrow \deg(\text{node}_j) + 1$
11:         **end if**
12:     **end for**
13: **end for**

It is well known that the distribution of the degree of nodes follows the power-law, that means a few nodes in *V* occupy a large portion of the edges in *E*. Networks such as WWW and online communities [17], cellular networks in biology [18], patients of sexually transmitted diseases [19], etc. have been known to follow the power-law degree distribution. We basically regard the generation process above of SFN as a model capturing real social behaviors of people approximately, comparing with other existing models. For example, all the nodes are connected to the same number of edges in a regular graph, which is inconsistent with the real inequality of peoples' social activities in the real world. The small-world networks [20] have been used as a model to capture the contacts of people where local communities and their weakly bridging ties play each role in the infection spreading [21, 22]. However, let us not choose this model because we embrace an aim to clarify the effect of contacts with external communities caused by connections due to unintendedly receiving contacts from others such as the thin $W$-$m_0+1$ edges in Fig. 1. An individual in the real society may not distinguish between a bridge between communities and other relationships, but can distinguish a relationship one intentionally made with others and unintended relationships made by others. This distinction shall enable an individual to choose edges if intended relationships are more choice-worthy. In addition, a small-world network e.g. in [20] starts its own generation from a regular graph and moves a certain number of randomly selected edges connected from each node, which does not match with humans' behavior intending to create a community in the real world. In the more recently proposed Mediation-Driven Attachment Model [23, 24], each new node first picks an existing node at random and connects not directly with this but with *a certain number* of its neighbors also picked at random. This is an extension of SFN supposed to result in having each new node connected to "rich" people linked to a large number of "poor" people of low degrees by choosing link destination nodes by the probability estimated to be inverse of the harmonic mean (IHM). Although this may make the network fit to the real society, we do not choose as the model for this paper because the attendants of places do not have a bias to low-degree nodes in the daily life human behaviors. Furthermore, the network of locations (i.e. rooms in this paper) has been shown to form a scale-free network [21], and the networks we deal with model the mixture of people and rooms. In this sense, the idea to borrow SFN as a backbone is at least partially supported as a tool for modeling a real-space human society in this paper. Thus, considering the suitability of each model to the intuitive understanding of the social life of people, we take SFN as the backbone and revise it with introducing constraints corresponding to the spatiotemporal restrictions in the real life of people as in the next section. Reference [19] is an example supporting this choice in the sense a virus infection network has been shown to follow the power-law distribution. We can find other literature utilizing SFN as a model for the analysis of infection dynamics [25].

## Constrained scale-free networks considering spatiotemporal restrictions

The SFN may be, as discussed above, regarded as a natural model for capturing human social behaviors as long as there is no restriction on the reach of social behaviors of each human. On the other hand, the physical constraints in the real living environment restrict the width of the room where people may gather, and the time - a person cannot meet as many people as one would like to meet. Thus, the original scale-free network cannot be used as it is to model real social behaviors of people. In this sense, let us revise the process as follows. Besides, we need to deal with not only humans but also spaces of rooms as nodes assuming that these are working as entities attracting people and that these entities are dealt with to meet the requirements of humans. Furthermore, the virus is wrapped in saliva and cast into the air in meeting rooms. We reflect the spatiotemporal restrictions as constraints in the two types of artificial networks of people and things/places.

The first artificial network we use in this paper is SFN with Spatiotemporal Constraints (Algorithm 2: SFN-SC). A feature of SFN-SC is that it starts from multiple cliques rather than a single clique used in the sheer original SFN shown above. This means we can assume multiple ($K>1$) groups of people that may attract people who may come later, which is a generalization of the original SFN that corresponds to setting $K$ to 1. A spatiotemporal constraint is given by the next feature. That is, the degree of the link destination node (the partner of the new connection) from the new node is constrained by the given upper bound of $W$ at each time a new node is added, as in line 10 of the Algorithm 2. If the destination node means a room, the room capacity is at most $W$ people for the protection of COVID-19 infection spreading. The $m_0$ nodes to which this room directs via its $m_0$ out-going edges as in lines 10 through 13 may correspond to the facilitators (or attracting speakers) in the meeting. These facilitators are people occupying a part of the capacity $W$ but do not join in the group of the other people in the room who are sheer participants. The facilitators direct their out-going edges to other nodes, which means to inform about (e.g., call for attention or report the achievements of) the meeting. If the destination node means a human, other nodes coming after to be connected to the node mean people who choose to meet the person. Because these aftercoming nodes may desire to meet in the same space at the same time, in both cases, i.e., where the destination node is a room or a human, we add the lines from 17 through 24 for showing a group can be made by the aftercoming nodes which may generate an infection environment.

**Algorithm 2: SFN with Spatiotemporal Constraints (SFN-SC)**
1: $E \leftarrow \{\}$.
2: $V \leftarrow \bigcup_{k=1}^{K} V_k$ **where** $V_k = \{\text{node}_{(k-1)\min(m0, W)+1}, \text{node}_{(k-1)\min(m0, W)+2}, ..., \text{node}_{k\min(m0, W)}\}$
3: **for** $k = 1 : K$ **do**
4: $\quad E \leftarrow E + \{\text{edge}_{i,j}\}$ **for all** $\text{node}_i \in V_k$, $\text{node}_j \in V_k$ $(i \neq j)$
5: **for** $k = 1 : K$ **do**
6: $\quad G = E + \bigcup_{k=1}^{K} V_k$
7: **for** $i = m_0+1 : N$ **do**
8: $\quad$ **add** $\text{node}_i$ to $V$
9: $\quad$ **for** $j = m+1 : N$ $(i \neq j)$ **do**
10: $\quad\quad$ **if** $\deg(\text{node}_j) < W$ **and** $rank_{node_j \in V} \text{random}(1) \deg(\text{node}_j) \leq m_0$
11: $\quad\quad\quad E \leftarrow E + \{\text{edge}_{i,j}\}$
12: $\quad\quad\quad \deg(\text{node}_i) \leftarrow \deg(\text{node}_i) + 1$

```
13:                    deg(node_j) ← deg(node_j) + 1
14:              end if
15:         end for
16: end for
17: for i = m+1 : N do
18:      group_i ← {}
19:      for j = i+1 : N   (i ≠ j) do
20:           if edge_{i,j} ∈ E
21:                group_i ← group_i ∪ {node_j}
22:           end if
23:      end for
24: end for
```

Thus, the two constraints below are assigned on each node X in SFN-SC:

Constraint 1: From other nodes, X accepts at most W edges including the edges X directed to other nodes that appeared in G before X, as shown by $node_j \in V$ in line 10. This means X accepts at most W-$m_0$ nodes from newer nodes i.e., nodes that appear in G after X.

Constraint 2: X directs an edge only to nodes that joined G before X, as $node_j \in V$ in line 10.

By the combination of these two constraints, the degree of each node X is bounded by W. X receives edges from newer nodes that appear after X only when deg(X) is less than W. If X directed edges to other $m_0$ nodes, it accepts additional edges only if W > $m_0$. On the other hand, if $W \leq m_0$, nodes newer than the $K \min(W, m_0)$ i.e., $KW$ nodes in the initial clusters generated in line 2 cannot get connected to those clusters but generate new clusters that grow to include W + 1 nodes.

However, it does not fit to our daily experience in urban life to assume that each habitant contacts others spontaneously only when one newly joins the city, or that one chooses to contact only people who lived since before oneself. In this sense, we create the other procedure in Algorithm 3 where Condition 2 is expired ($node_j \in V$ in line 9 no more means constraint 2, because V includes all the N nodes to be included in G from the beginning of this algorithm due to the initial members of V given in line 1). Due to the constraint releasing, the degree of each node X is not strictly bounded by W i.e. X can choose other nodes to contact after it receives edges here. We call this algorithm of network generation "selfish" because one (node X) does not accept edges from other nodes if X is busy for the contacts with W other nodes even if X is responsible partially (i.e., the W nodes may include ones to which X directed edges for his own interest), but X executes his right to direct new edges to $m_0$ other nodes even in the busy situation. This matches the real-world human society, where a busy person protects oneself from new contacts but keeps calling others for one's own business. It is noteworthy that X denoted by $node_i$ in Algorithm 3 may contact other $node_j$ where j is larger than i, which means someone who does not belong to V before $node_i$. This means X may endeavor to contact the potential market embracing new people who are not yet popular if X has room within his edges ($m_0$), because X contacts not only those who chose to contact others before X does but also who do so after X. As in Figures 2 and 3, Algorithm 3 expands the connection from a cluster to other nodes (see the connections from node 10). The degree distribution of neither SFN-SC nor SFN-SSC follows the power-law but the most popular nodes come to take the degree W or W+$m_0$. Thus, we mean to highlight the effect of unintended connections (represented by W− $m_0$) by controlling W and $m_0$, rather than simulating the spread of infections in the original SFN.

**Algorithm 3: SFN with Selfish Spatiotemporal Constraints (SFN-SSC)**
1: $E \leftarrow \{\}$ ; $V \leftarrow \{\text{node}_1, \text{node}_2, \ldots, \text{node}_N\}$ ; $G = E + V$
2: **for** $k = 1 : K$ **do**
3:    $V_k = \{\text{node}_{(k-1)\min(m0, W)+1}, \text{node}_{(k-1)\min(m0, W)+2}, \ldots, \text{node}_{k\min(m0, W)}\}$
4:    $E \leftarrow E + \{\text{edge}_{i,j}\}$ **for all** $\text{node}_i \in V_k$, $\text{node}_j \in V_k$ $(i \neq j)$
5: **for** $i = m_0 + 1 : N$ **do**
6:    $\deg(\text{node}_i) \leftarrow m_0$
8:    **for** $j = m+1 : N$ $(i \neq j)$ **do**
9:       **if** $\deg(\text{node}_j) < W$ **and** $rank_{node_j \in V} \text{random}(1) \deg(\text{node}_j) \leq m_0$
10:         $E \leftarrow E + \{\text{edge}_{i,j}\}$
13:         $\deg(\text{node}_j) \leftarrow \deg(\text{node}_j) + 1$
14:       **end if**
15:    **end for**
16: **end for**
17: **for** $i = m+1 : N$ **do**
18:    $\text{group}_i \leftarrow \{\}$
19:    **for** $j = i+1 : N$ $(i \neq j)$ **do**
20:       **if** $\text{edge}_{i,j} \in E$
21:         $\text{group}_i \leftarrow \text{group}_i \cup \{\text{node}_j\}$
22:       **end if**
23:    **end for**
24: **end for**

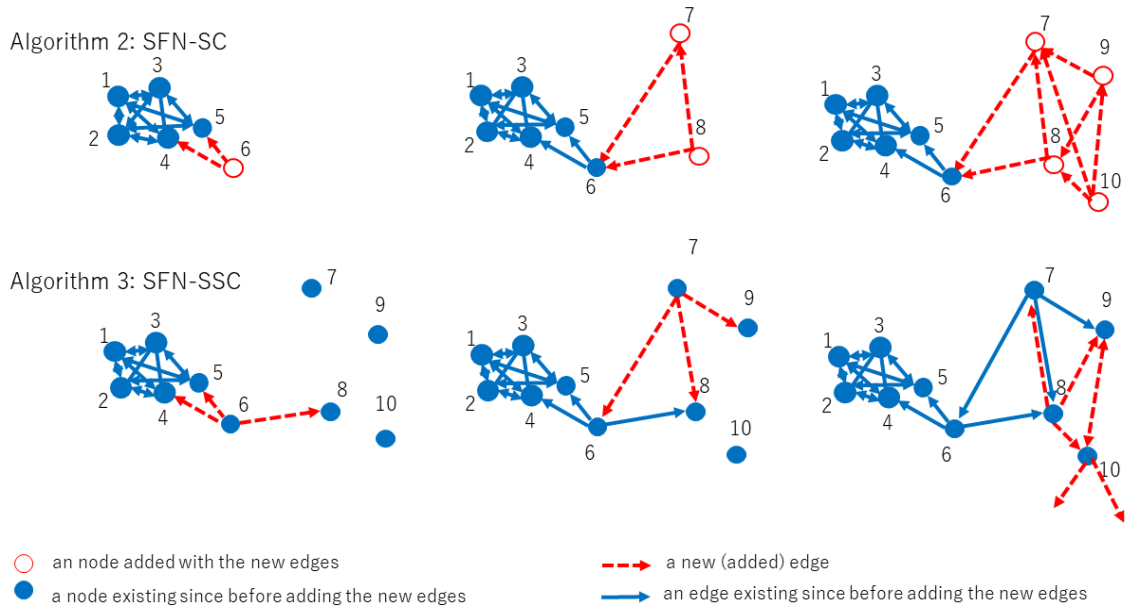

**Fig 2. The intuitive comparison of Algorithms 2 and 3: The red (hollow) nodes and dotted lines show new nodes and edges, the blue solid lines existed before the red. SFN-SSC makes a wider connection between communities.**

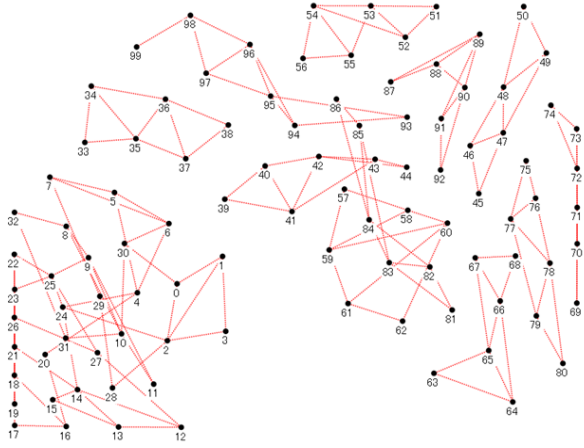 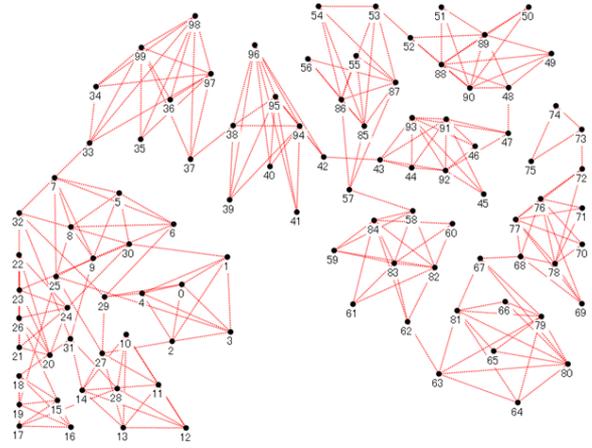

**Fig 3. An example of obtained graphs by the two algorithms. SFN-SSC connects clusters although the setting of the number of people each member accepts to meet ($W$) and the number one chooses to meet ($m_0$) are set equally for the two algorithms.**

## The model of infection spreading dynamics

For each node in the network generated above, the three-step dynamics is considered here for spreading the infection. The first is to catch a virus from neighboring nodes (i.e., humans, places, or things in the places), the second is to get infected, and the third is to become an infector to other nodes. Here note each edge in $G$ represents the possibility that two entities (humans, places, or things in the places) may contact. An activation rate $p$ is defined here as the percentage of "active" edges and groups (generated by lines 17 through 24 in Algorithms 2 or 3) connecting the nodes that contact in the simulation. That is, $p|E|$ edges are used for the nodes' contacts and all the other edges are shut off. And here, a contact of nodes means the nodes meet closely enough, e.g. a conversation of 15 minutes at the distance of 2m, to cause an infection if one of the nodes has infectivity. If people contact fully activating all the edges in $G$, $p$ is 1.0. If the frequency of contacts is reduced by 80% due to governmental regulation, $p$ is set to 0.2. These steps are supposed here to occur as follows, week by week:

*Contacting the neighbors*: the rate of $p$ (a real value from 0 to 1) of all the edges in $E$ and of all the groups organized in the Algorithm 2 or 3 are randomly chosen. Then, for each node $X$, the other nodes connected to $X$ via an active edge or are in the same active group as $X$ are taken as the contacting neighbors of $X$.

*Catching virus from neighbors*: catch($i$) takes the maximum value of infector(neighbor) of all the contacting neighbors of node$_i$. In other words, a node catches virus if any of its contacting neighbors has become an infector. An infector is generated on the following rule.

*Getting infected*: node$_i$ is infected if catch($i$) is larger than a random value which ranges by uniform probability between 0 and 1, meaning one gets infected by the probability of 0.5 if one catches viruses from any neighbors. The strength of infection is here given by a real value infected($i$) rather than a discrete judgment considering the uncertainty. The infection of a node fades by a constant $r$ (set to 0.7 corresponding to the recovery in 4 weeks) each week. This fading means the inactivation of viruses or the node's acquisition of immunity.

*Becoming an infector*: If a node is infected, one infects others that is represented by the value of infector(*i*) equal to 1 by the probability of 0.2 on the announcement of WHO on April 10, 2020. Otherwise, infector(*i*) is equal to 0. infector(*i*) also fades by *r* per week.

Suppose here that anyone linked to other *n* people by the probability of *p* is expected to infect others to generate approximately two other infected nodes (see https://rt-live-japan.com/ for the data on effective reproduction numbers) for the two weeks. Here, an individual gets infected by 50% in the setting above if one catches viruses, and infects others by the probability of 0.2. Thus *n* \* *p* \* 2 \* 0.2 \*0.5 is supposed to be nearly equal to 2.0, which means *p* is approximated by 1.0 in the usual daily-life communications if *n* = 10. Here, 10 is the approximated average number of friends reported for each person in Japan (https://chosa.nifty.com/relation/chosa_report_A20121123/1/). The effective reproduction number 2.0 is larger than the reported number of infection reproductions in the period of spreading [26], but set to this value to consider the situation people communicate as usual. Infection dynamics have been also modeled by fusing a network-based and the population-based SIR model [9:1, 25]. However, we take the infection process above in order to model the probabilistic threshold system where each node *X* gets infected if catch(*X*) exceeds a threshold and then infect others by a certain probability.

# Results

We executed the simulation of 100 weeks/trial setting the number of nodes (*N*) in *G* to 1000 and 10000, and two starting infected nodes in the 0-th week, one in a cluster (meaning to have degree *W*) and another out of any cluster.

## The tendencies of infection (1): the second waves

Figures 4 and 5 show the average number of new infection cases per week on Algorithm 2 and 3 for 10 trials. Note each curve shows not a history in one trial but a mixture of the sequences of the averaged trials. The number of initial clusters is set to **$K = 0.1$ *N/W*** in order to make separate clusters for a large *W*. For example, if the number of nodes (*N*) is 10000 and *W* is 20, we start from 50 cliques as initial clusters. In these figures and after, cases(*t*) means the number of new infection cases in the *t*-th week.

As in Figures 4 and 5, we find several tendencies. First, the peaks of the second wave tend to be lower than of the first wave (e.g., the third of Figure 4(a), the first of Figure 4 (b), the second and the third of Figure 5 (a), and the first of Figure 5(b)) as far as the structure of the social network is invariant throughout the simulated period. We detected the second peak in the 190 curves (39.6%) among the 480 cases (twenty values of *W*, three values $m_0$, single and multiple initial cliques, two values 1000 and 10000 of *N*, for the two Algorithms) we tested, of which 19 (3.9%) was higher than the first peak. Only 2 (0.4%) of curves had the second peak after going down to less than 10% of the first peak, in both of which the first peak was less than 10 infected cases. Thus, no second peak in the severe cases exceeded the first. Here, a peak means the time *t* where $f(t) = \max_{t-\Delta t \leq s \leq t+\Delta t} f(s)$ setting $\Delta t$ to 5 weeks.

In the next place, we changed the network at the 20th week from a multiple to a single initial clique on the way of infection spreading that is an imaginary change without changing *N*, the density of the network, or the infection status (infected(*X*), infector(*X*), catch(*X*) defined in the model of infection) of any node. This means an experiment of Scenario C in Figure 1, i.e., to change the structure of the entire graph *G*. As a result, as in Figure 6, we found the second finding: We detected the second peak in the 54 curves (67.5%) among the 80 tested cases (eight values of W, five values of $m_0$, multiple initial cliques, two values 1000 and 10000 of $N$, for the Algorithm 3 because here we aim to investigate the changes in the inter-cluster structure that is richer in SFN-SSC) we tested, of which 21 (25.1%) was higher than the first peak. Furthermore, 11 (13.1%) of curves had the second peak following the structural change, after going down to less than 5% of the first peak. However, as in the case of $(W, m_0)=(8, 4)$ in Figure 4(b), the simulation using SFN-SC can result in the meaningless small number of new infection cases. This happens because an initially infected node comes to be located in a small cluster isolated from the other nodes in $G$ due to the feature of SFN-SC as shown in Figure 2. For this reason and also because SFN-SSC reflects the above mentioned "selfish" nature of humans in contacting others, we take the exemplified curves from the results of SFN-SSC.

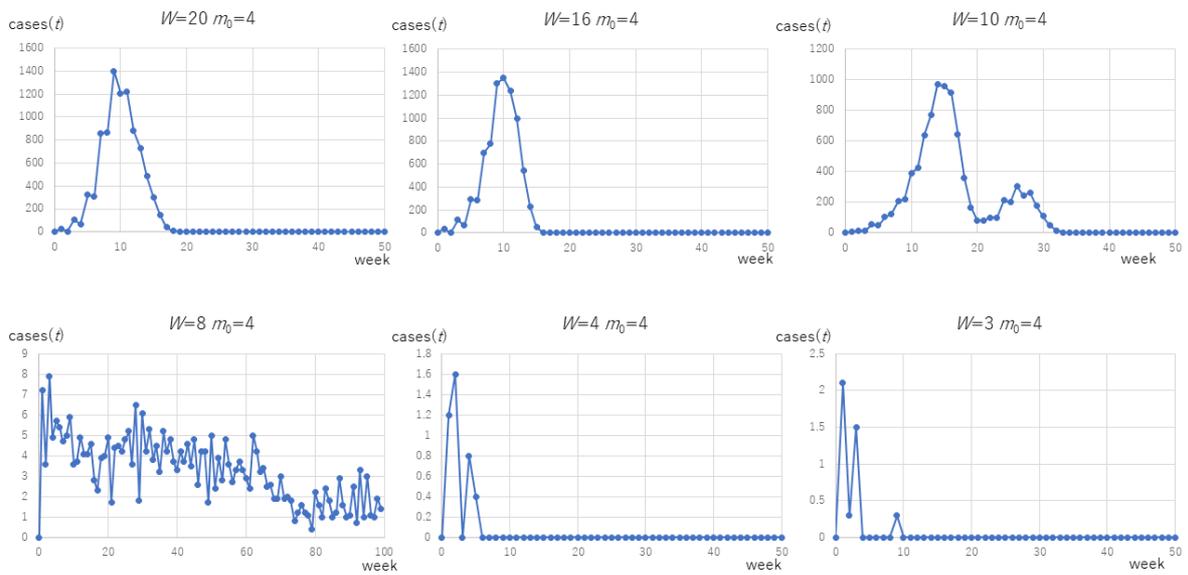

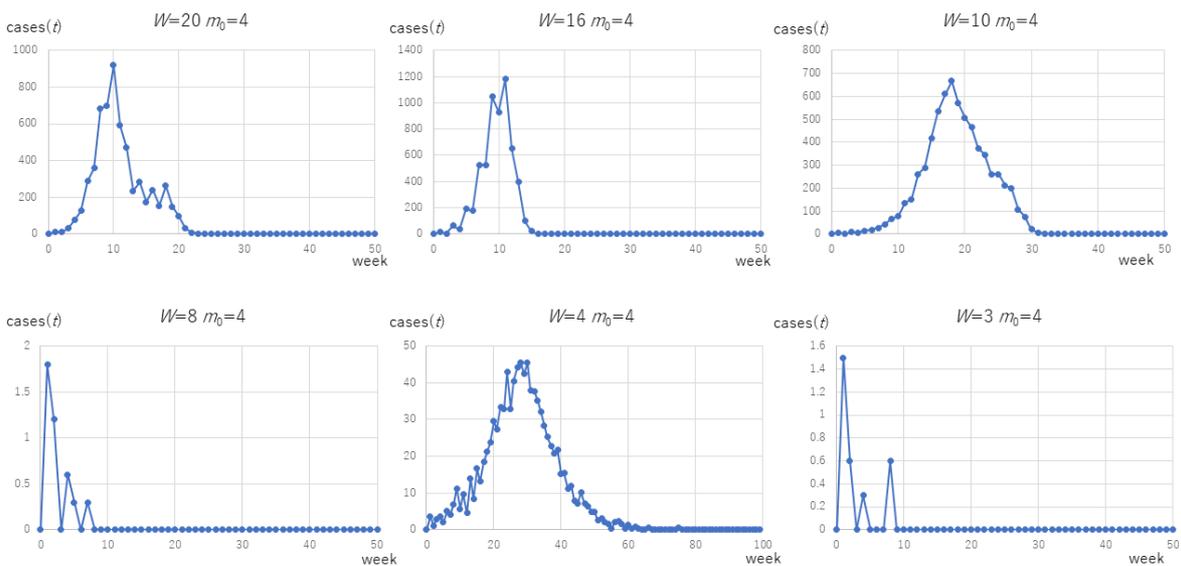

**Fig 4. The number of new infection cases (a) $K=1$ and (b) $K = 0.1$ $N/W$ of SFN-SC (Algorithm 2, $N = 10000$) for varying $W$ and $m_0$.**

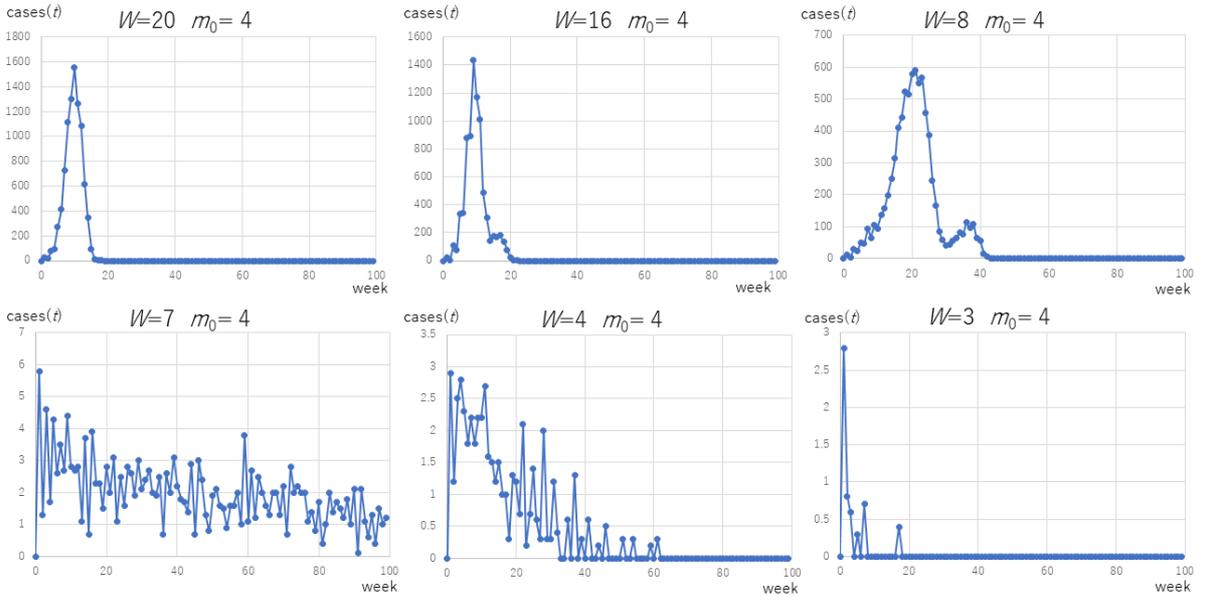

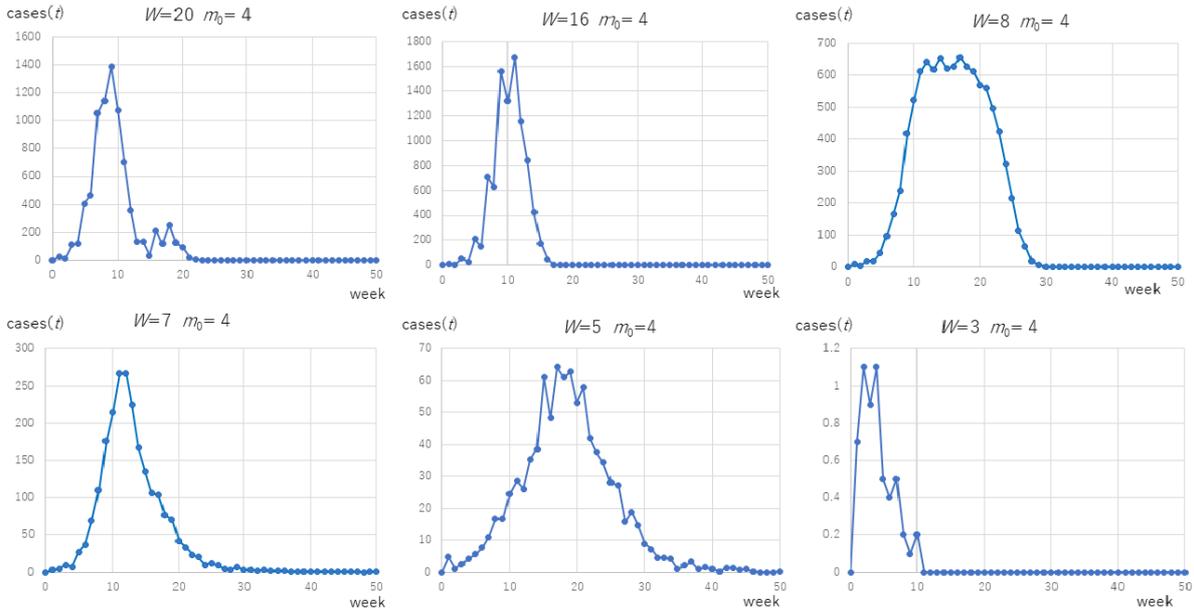

**Fig 5. The number of new infection cases (a) $K=1$ and (b) $K = 0.1$ $N/W$ of SFN-SSC (Algorithm 3, $N = 10000$) for varying $W$ and $m_0$.**

## The tendencies of infections (2): for varying $W$ and $m_0$

The **third** tendency we find from the curves exemplified in Figs 4 and 5 is that the peak of the number of infection cases tends to vary for various values of $W$ and $m_0$. In Tables 1 and 2, we recognize two tendencies in the range of large $W$ and $m_0$. First, the number of infection cases depends positively on $W$ according to the values in the cells in each column. Second, the larger value of $m_0$ does not always result in a larger number of new infection cases. For $W$ of 6 or larger, the largest number of infections occur for $2 \leq m_0 \leq 8$ rather than for a larger $m_0$. In the cells where $W < 6$ or $m_0 < 2$ (especially for less than 2), the number of new infection cases are substantially smaller than in the other part of the table. More importantly, the maximum number of new cases rises radically from $W$ less than $m_0$ to larger than or equal to $2m_0$. Focusing on the range of $2 \leq m_0 \leq 8$, let us see the tendency as in Figure 7. For the cases of multiple initial cliques ($K = 0.1\ N/W$), two-step growth i.e., at $W = m_0$ and $W = 2m_0$ is observed. The growth at $W = 2m_0$ is especially obvious for the case of a single initial clique ($K = 1$).

In the Tables 3 and 4, the average timing of infection for all the cases is shown. These values mean the expected timing of infections, setting the first infections in this network as the 0-th week. In this table, the third tendency is observed: the average infection week tends to rise radically from $W$ less than $m_0$ to larger than or equal to $2m_0$, and is reduced for the larger $W$.

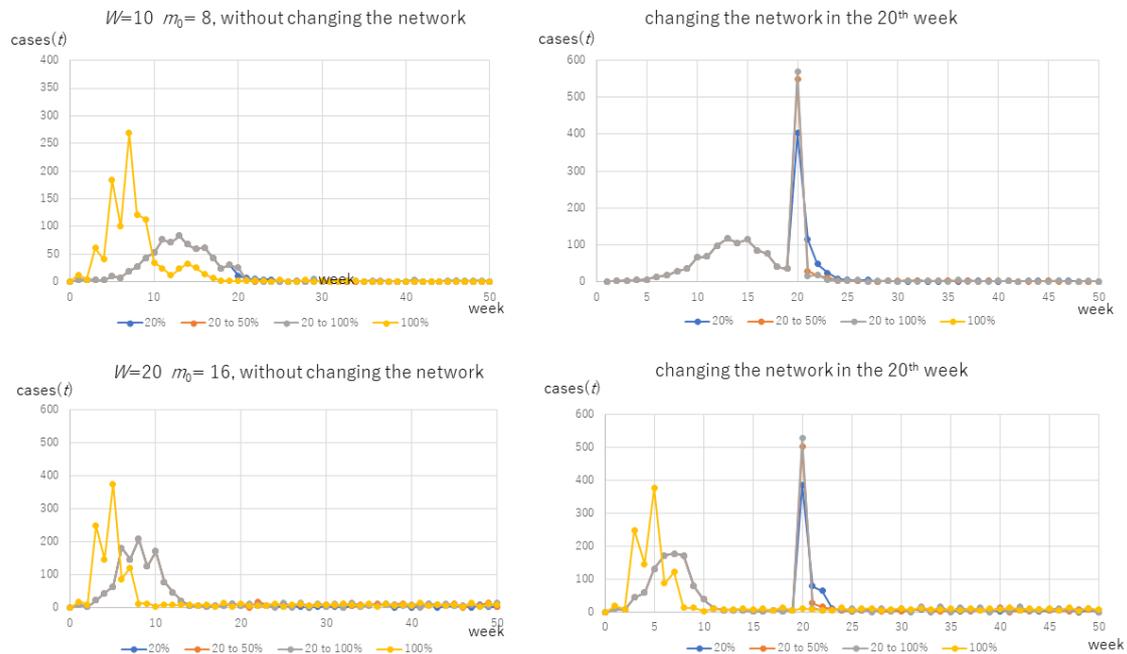

**Fig 6. The effect of changing the network structure: the effect of Scenario C in Figure 1.**

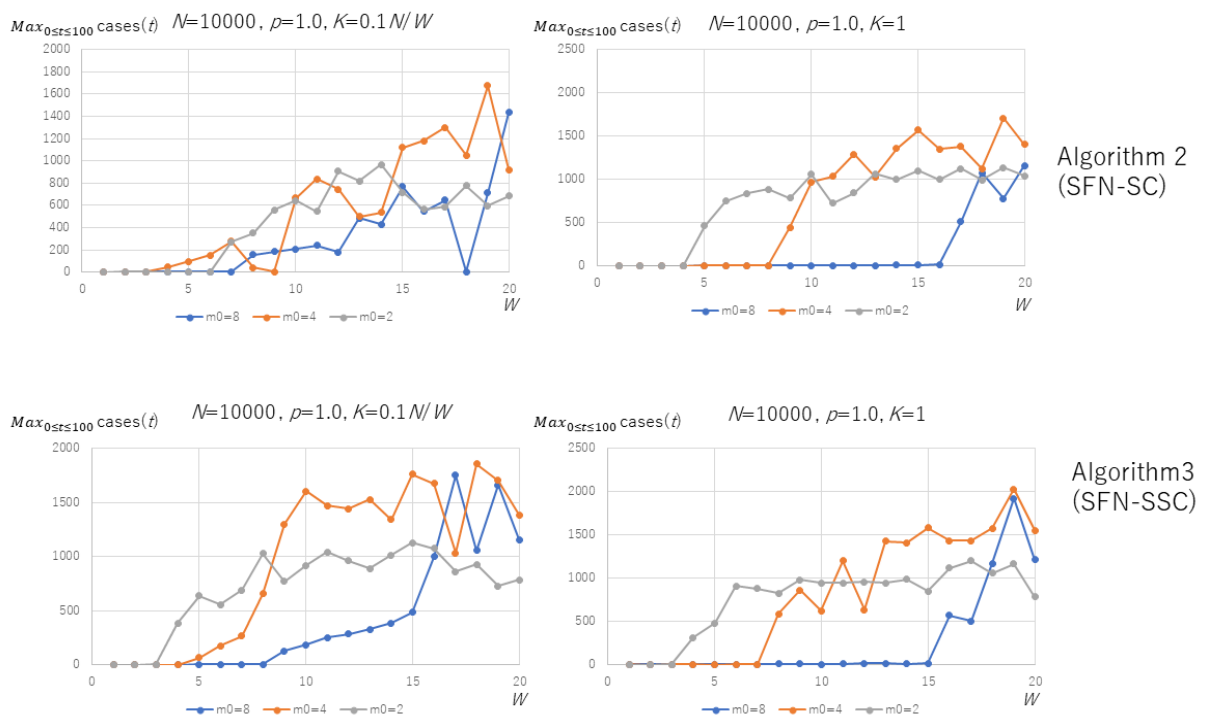

**Fig 7. The number of new infection cases for $K=1$ and $K = 0.1\ N/W$ of SFN-SC and SFN-SSC, $N = 10000$, varying $W$ and $m_0$.**

## The tendencies of infections (3): the releasing of restrictions

In Figure 8, the examples of releasing restrictions of 20% ($p = 0.2$) to 50 or 100% (completely free communication) are shown. The restrictions are released at the 20th week in all the 200 cases including the five trials of eight values (32, 20, 16, 10, 8, 4, 2, 1) of $W$, five values of (16, 8, 4, 2, 1) of $m_0$. Among these cases, we took the 55 cases where the peak of 20% was before the 20$^{th}$ week and evaluated the correlation of $X$: the closeness of the original (i.e. $p = 0.2$) highest peak to the 20$^{th}$ week and $Y$: the height of released (i.e. $p = 1.0$) after-wave relative to the original highest peak. As a result, a strong correlation is shown between $X$ and $Y$ (Pearson's correlation was 0.81) as in Fig 9. $X$ and $Y$ are respectively defined below:

$$X := \text{Max}_{p=0.2,\ 20 \leq t \leq 100}\ \text{cases}(t) / \text{Max}_{p=0.2,\ 0 \leq t \leq 100}\ \text{cases}(t) \quad (1)$$

$$Y := \text{Max}_{p=1.0,\ 20 \leq t \leq 100}\ \text{cases}(t) / \text{Max}_{p=0.2,\ 0 \leq t \leq 100}\ \text{cases}(t) \quad (2)$$

In Eq.(1), $\text{Max}_{p=0.2,\ 0 \leq t \leq 100}\ \text{cases}(t)$ and $\text{Max}_{p=0.2,\ 20 \leq t \leq 100}\ \text{cases}(t)$ respectively mean the largest cases($t$) for all the simulated 100 weeks and for the period after the release in the 20$^{th}$ week, both setting $p$ to 0.2. The larger $X$ means the closeness of the 20$^{th}$ to the highest peak before the 20$^{th}$ week because $X$ is reduced to be the less if the 20$^{th}$ week appears the later after the peak. Note $X$ is not the sheer temporal distance but the relative delay considering the decrease in the number of infection cases reflecting our aim to investigate the influence of time reducing cases($t$). On the other hand, in Eq.(2), $\text{Max}_{p=1.0,\ 20 \leq t \leq 100}\ \text{cases}(t)$ means the largest cases($t$) after the release in the 20$^{th}$ week. The larger $Y$ means the stronger impact of the release because Y approaches 1.0 from a smaller value if the value of $\text{Max}_{p=1.0,\ 20 \leq t \leq 100}\ \text{cases}(t)$ i.e. the largest number of infection cases after the release, is equal to or larger than the largest number before the release. Thus, the result here means that the sooner is the release after the peak in a restricted period, the higher wave is caused after the release.

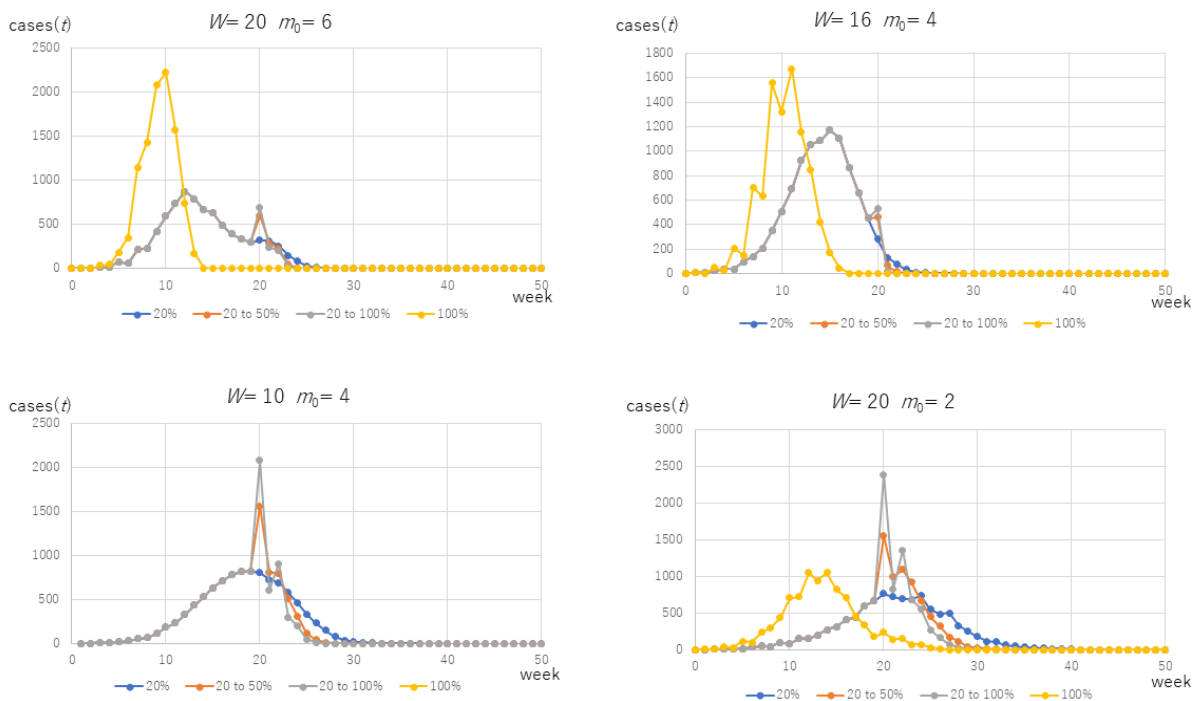

**Fig 8. The number of new infection cases before and after releasing the restriction for SFN-SSC, $N=10000$.**

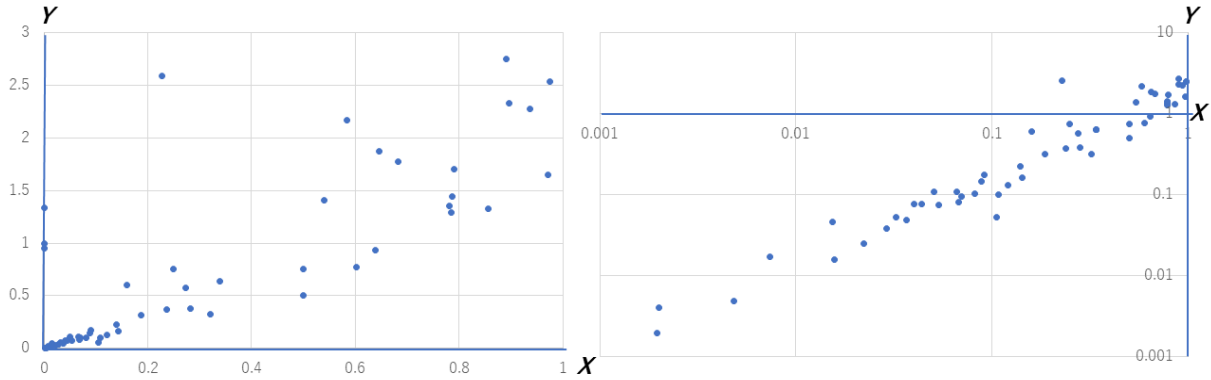

X: The closeness of the original highest peak to the 20th week (releasing time)
Y: The height of the released after-wave relative to the original highest peak

**Fig 9. The impact of releasing the restriction early.** *X*: the closeness of the original (i.e. *p* = 0.2) highest peak to the 20th week and *Y*: the height of released (i.e. *p* = 1.0) after-wave relative to the original highest peak.

## Discussion

The first finding that the peak of the second wave tends to be lower than the first, if the network structure is invariant, can be explained as Scenario A in Fig 1, where the infection cases monotonously fade after the peak, and as Scenario B, where the second wave occurs in a localized cluster in the network. Unless the localized clusters are connected, they get infected from the initial cluster independently and fades without and interactive activation. Thus, this first finding can be explained as the dynamics of a network with distributed local clusters with weak ties.

The second finding is that the peak of the second wave could be higher than the first if the network changes the structure on the way of infection spreading or after its suppression. This occurs because the infectors come to directly touch non-infected people who have not acquired immunity. And this implies a risk of an outbreak even at a time when the observed infection spread is settled if people activate inter-community interactions.

The third finding is that the number of infection cases depends positively on *W*, the largest allowed number of people one accepts to meet. On the other hand, $m_0$ for the largest number of infection cases tends to take an intermediate value (such as 2, 4, or 8) rather than the larger value (such as 16). The author's hypothesis for explaining this result is that the people one meets tend not to have acquired immunity yet if the number of people to meet is restricted to a narrow community and may cause infections to spread. This lack in immunity is supposed to occur in the sparsely connected parts of the entire graph *G* because people on the nodes in such parts are of lower degree and fewer opportunities to touch others. If there are large meeting groups the hypothesis goes that the infections in these groups attack the nonimmune part of the network. If $m_0$ is constrained to 1 or 0 (meaning to choose nobody to meet), according to Table 1, the risk is reduced significantly because the casting and catching of viruses occur in lower probability. However, this is a too strong constraint for an ordinary way of living.

Thus, a realistic way we obtain from the third finding may be to reduce the participants in a group (i.e., $W-m_0$). As in the results of Figures 7 (and Table 1), reducing $W-m_0$ to 0 contributes to suppressing the number of infected cases. The reason for this is clear: if W is equal to $m_0$ or smaller, no new node can be connected to a cluster where all the nodes have a degree of $m_0$ or larger in both Algorithms 2 and 3. In such a case, no cluster can be larger and the spreading is restricted to each small cluster. However, restricting each group size to 0 means having no group, that is another hard constraint for businesses especially if the businesses need to create contracts with new actors. Therefore, we should pay attention to the other threshold $W=2m_0$, as in the results in Figures 7 for the two types of networks generated from both single and multiple initial cliques. The condition $W \geq 2m_0$ causes radically high peaks, which can be explained by considering the two-step phenomenon caused as follows.

Step 1) A node X is added to G with $m_0$ edges from X to existing nodes. As a result, $m_0$ as a part of the capacity of G to accept connections to new nodes via new edges is lost.

Step 2) X itself has a capacity to accept $W - m_0$ edges, which comes to be added to the capacity of G. Thus, if $W - m_0 \geq m_0$, i.e., if $W \geq 2m_0$, each cluster in G obtains the more capacity with adding new nodes. Otherwise, each cluster stops the growth by a limited size because each cluster in G loses its capacity to get new connections.

Thus, we propose each node e.g., person or a room, accepts less than $2m_0$ i.e., twice as many people to touch as the people the node chooses spontaneously to meet. More easily put, it is safe if each individual do not accept others whom one does not choose (keep $W < m_0$) and, if this is too strict, one should not accept contacts from a larger number of people as one chooses to contact (keep $W < 2m_0$). One way to understand $W < 2m_0$ is to rewrite this as $W - m_0 < m_0$. Because $W - m_0$ means the number of other people one accepts contacts from without one's own choice, this means one should keep in one's own community that one chose intentionally rather than accepting new contacts from external communities. This is conceptually common to the finding in the literature [1, 27] that mixing heterogeneous people or connecting nodes via a long-distance edge can result in infection expansion. The effect of a bridge between communities is just one interpretation of $W - m_0$ and $W - 2m_0$ under the assumption that directing edges to $m_0$ other nodes is an intentional action from a node. However, the results of the timing of infection (Tables 3 and 4) add evidence of the creation of bridges between clusters of nodes. That is, the clusters start to be bridged in the range of $m_0 \leq W < 2m_0$ that enables infections through long paths in G taking a longer time than in each small cluster. For a larger W, the paths are widened due to the added edges and the infections are sped up to reduce the time for spreading.

Finally, the fourth tendency we found is that the release of once given constraint may trigger a second wave higher than the peak of 100%, i.e., than the spreading without introducing any constraint from the beginning if the release is introduced soon after the peak. Thus, we can say the restriction, if introduced once, should not be released quickly without checking the results of the step-wise release of restrictions. For example, the government should release from p=0.2 to 0.3 without allowing public places where people from different communities often meet (this comes from the second finding), see what happens to the number of infection cases, and go ahead to 0.4 only if a substantial increase is not found.

We find relevances of our results to previous work in the literature. The results in this paper can be associated with the assertion that there is a risk of infection spreading in a society where people in each household decide to maintain an in-person social connection with one person from other households [28]. In comparison, the suggested policies above are not so strict but realistic and practical proposals for the decision making of both individuals and policymakers. In [29], by combining a stochastic model for state transitions and networked multi-agent simulations, they showed the possibility that undetected (asymptomatic or mildly ill) cases may still exist and cause complex social influence even after the peak of the pandemic wave. In our paper, the asymptomatic cases correspond to nodes infected but not chosen probabilistically to be infectors before they lose the infectivity. These nodes are not so densely surrounded by infected nodes as infectors are, because infectors can infect neighbors. Therefore, the risk here is relevant to the case of second waves in Scenario B. In [30], the slowing down of spreading has been found to be caused by weight-topology correlations. The weight here means the frequency of communications via each edge, whereas the topology corresponds to the location of each edge i.e. whether it is in a cluster or a bridge. Interpreting that the frequent communication across bridges between communities accelerate the spreading of infections, we are verifying similar tendencies in different setting. An algorithm for reconstructing the strong ties that are the invariant backbone structure of a network, divided from weak ties made of transient links in the dynamic network structure, from data on the individuals' states in a spreading process [31], has been shown to be useful for a targeted immunization strategy that prioritizes influential nodes in the inferred backbone. Our conclusion may seem to differ from this result in that we showed the role of bridges between clusters on the spreading, that seem to correspond to weak ties between strong ties. However, the bridges between clusters we dealt here are not necessarily as transient as weak ties and are really influential to various clusters, so is consistent with [31].

## Conclusion

The infection spreading has been simulated over human networks where the constraints are given by the maximum number of people that each person accepts to meet ($W$) and the number of other people one chooses to meet ($m_0$). The real-space urban life of people is modeled as a modified scale-free network with constraints on the constants on the two width factors $W$ and $m_0$ above. As a result, four findings have been obtained as in the discussion, that we here put within this one paragraph. That is, for suppressing the number of infections, we need policies with attention to the bridges between densely infected clusters rather than just looking at the clusters. Simply put as recommendations, the government should release once introduced restrictions on peoples' communications late enough, with careful attention to the result of step-by-step releasing. And for individual people, it is safe if each individual can avoid contact from others whom one did not choose. This means staying with ones' own intimate community if one chooses to meet intimate people. Even if this constraint is too strict, one should not accept additional contacts from as large number of people as one chose to contact during the period of infection spread. It should be noted here the reader should not misunderstand that one can meet as many people as one likes as far as the number of additional contacts is less than the chosen contacts. The point is instead that one should look back at people one chose to contact so far during the period of infection spread and then restrict the new contacts following the rule above because the events in the network above have been simulated for this period.

In this paper, we related infection spreading to the bridges the changes in the structure of the social network. Here we intentionally restricted the dynamics of the network to just one-time change for clarifying the effect of a change. This is realistic for the release of $p$ to a larger value because the release of governmental restriction is usually one or a few times in a country or an administrative district. However, we consider our future step about continuous change in the network structure because the real society changes continuously once people in the public are allowed to interact with others freely. Hence our new target of research on temporal networks, to enable extended analysis for network epidemiology by referring to real data [32], via collecting essential datasets using methods for finding and inventing useful data [33].

# Acknowledgment

# Funding

This study is in progress, partially supported by grant JSPS Kakenhi 19H05577 and a partial extension of our studies in JSPS Kakenhi 16H01836.

**Table 1. The peak number of infection cases per week for 100 weeks, $p = 1.0$ (the cells in denser red color show the larger values in each column)**

$K = 1: N = 1000, p = 1.0$

| W \ $m_0$ | 1 | 2 | 4 | 8 | 16 |
|---|---|---|---|---|---|
| 32 | 27.9 | 115.8 | 209.1 | 225.5 | 32.4 |
| 20 | 9.5 | 155.9 | 135.5 | 181.3 | 34.1 |
| 16 | 19.5 | 114.3 | 113.3 | 12.7 | 6.4 |
| 10 | 5.6 | 95.8 | 107.5 | 10.9 | 7.8 |
| 8 | 7 | 74.6 | 9.2 | 4 | 7.2 |
| 6 | 4.8 | 84.1 | 3.6 | 2.4 | 5.4 |
| 4 | 3 | 3.2 | 2.8 | 5.6 | 3.8 |
| 2 | 2.4 | 1.2 | 1.4 | 1.4 | 4.5 |
| 1 | 0.9 | 1.4 | 0.7 | 0.9 | 3 |

$K = 1: N = 10000, p = 1.0$

| W \ $m_0$ | 1 | 2 | 4 | 8 | 16 |
|---|---|---|---|---|---|
| 32 | 46.5 | 1282.9 | 1470.3 | 1953.4 | 32.7 |
| 20 | 15.2 | 1061.5 | 1451.1 | 921.7 | 29.9 |
| 16 | 31.5 | 856.6 | 1151.8 | 27.2 | 8 |
| 10 | 18.9 | 1002.6 | 1037.7 | 12.4 | 11.2 |
| 8 | 13.3 | 747.6 | 10.9 | 3.2 | 7.2 |
| 6 | 14.5 | 602 | 5.1 | 3.6 | 4.8 |
| 4 | 3.2 | 2.8 | 1.6 | 5.2 | 8.4 |
| 2 | 1.7 | 1.2 | 1.4 | 3.2 | 6.8 |
| 1 | 0.6 | 0.9 | 1.8 | 2.7 | 6.7 |

$K = 0.1N/W: N = 1000, p = 1.0$

| W \ $m_0$ | 1 | 2 | 4 | 8 | 16 |
|---|---|---|---|---|---|
| 32 | 6.3 | 82 | 185.8 | 210.7 | 64 |
| 20 | 7.8 | 107.3 | 163.6 | 209.7 | 35.7 |
| 16 | 7.6 | 84.8 | 187.1 | 121.6 | 6.4 |
| 10 | 4.5 | 90.7 | 129.8 | 21 | 7.8 |
| 8 | 7.8 | 108.6 | 73.5 | 4.8 | 4.8 |
| 6 | 1.6 | 84.3 | 16.7 | 2.9 | 9.2 |
| 4 | 1.9 | 40.5 | 2 | 4.5 | 3 |
| 2 | 1.1 | 1.2 | 2.6 | 2 | 2.1 |
| 1 | 0.4 | 1 | 2.4 | 0.8 | 3.1 |

$K = 0.1N/W: N = 10000, p = 1.0$

| W \ $m_0$ | 1 | 2 | 4 | 8 | 16 |
|---|---|---|---|---|---|
| 32 | 28.2 | 1298.3 | 2279.5 | 2760.4 | 567.9 |
| 20 | 9.5 | 1050.4 | 2148.4 | 1887.2 | 375.7 |
| 16 | 9.2 | 1002.3 | 1259.2 | 773.4 | 11.2 |
| 10 | 15.4 | 779.1 | 1251.4 | 269.2 | 4 |
| 8 | 9.2 | 844.3 | 674.4 | 2.4 | 5.6 |
| 6 | 5.7 | 680.7 | 176.2 | 1.8 | 3.6 |
| 4 | 4 | 315.8 | 3.2 | 3.2 | 1.6 |
| 2 | 0.9 | 1.2 | 1 | 1.4 | 1 |
| 1 | 0.8 | 0.5 | 0.8 | 0.6 | 0.4 |

**Table 2. The peak number of infection cases per week for 100 weeks, $p = 0.5$ (the cells in denser red color show the larger values in each column).**

$K = 1: N = 1000, p = 0.5$

| W \ $m_0$ | 1 | 2 | 4 | 8 | 16 |
|---|---|---|---|---|---|
| 32 | 26.7 | 119.4 | 160.7 | 139.1 | 22.6 |
| 20 | 10.1 | 103.2 | 146.2 | 142.4 | 25.6 |
| 16 | 15.1 | 85.2 | 149.4 | 14.7 | 12.5 |
| 10 | 8.5 | 89 | 72.4 | 6.3 | 8 |
| 8 | 5.1 | 81.4 | 8 | 3.4 | 3.7 |
| 6 | 5.2 | 62 | 3.4 | 2.8 | 3.6 |
| 4 | 2.6 | 2.6 | 1.2 | 1.6 | 2.7 |
| 2 | 0.6 | 0.8 | 1.3 | 0.7 | 2.8 |
| 1 | 0.3 | 0.7 | 0.5 | 0.7 | 1.1 |

$K = 1: N = 10000, p = 0.5$

| W \ $m_0$ | 1 | 2 | 4 | 8 | 16 |
|---|---|---|---|---|---|
| 32 | 32.5 | 1008.3 | 1280.2 | 1444.1 | 42.6 |
| 20 | 17.4 | 900.2 | 1113.5 | 859 | 23.2 |
| 16 | 17.9 | 896.8 | 1080.7 | 15.6 | 8.8 |
| 10 | 18.2 | 875.1 | 717.6 | 5.9 | 4.7 |
| 8 | 8.2 | 530.9 | 6.9 | 5 | 3.5 |
| 6 | 13 | 568.6 | 2.1 | 3.5 | 3.4 |
| 4 | 5.6 | 2.4 | 1.9 | 5.4 | 4.9 |
| 2 | 1 | 1.2 | 1 | 2.2 | 2.7 |
| 1 | 0.5 | 0.8 | 1.4 | 1.3 | 2.4 |

$K = 0.1N/W: N = 1000, p = 0.5$

| W \ $m_0$ | 1 | 2 | 4 | 8 | 16 |
|---|---|---|---|---|---|
| 32 | 5.5 | 88.5 | 169.9 | 166.7 | 96.6 |
| 20 | 11.9 | 73.3 | 178.2 | 187.1 | 32.3 |
| 16 | 1.5 | 107.9 | 101 | 93.6 | 7.1 |
| 10 | 2.9 | 66.2 | 118.4 | 22.6 | 5.2 |
| 8 | 4.1 | 68.8 | 55 | 2.5 | 2.6 |
| 6 | 1.3 | 46.3 | 13.4 | 2.7 | 4.5 |
| 4 | 2.5 | 31.2 | 2.5 | 2.2 | 2.9 |
| 2 | 1 | 0.6 | 1.5 | 0.8 | 1.1 |
| 1 | 0.5 | 0.5 | 1 | 0.9 | 1.5 |

$K = 0.1N/W: N = 10000, p = 0.5$

| W \ $m_0$ | 1 | 2 | 4 | 8 | 16 |
|---|---|---|---|---|---|
| 32 | 28.2 | 631.3 | 1737.6 | 1486.2 | 625.9 |
| 20 | 15.7 | 1012.6 | 1113.7 | 1561.3 | 316.2 |
| 16 | 5.6 | 864.5 | 1571.3 | 631.7 | 11.9 |
| 10 | 11.8 | 883.1 | 1381.7 | 133.3 | 5.4 |
| 8 | 9.6 | 707.3 | 428.9 | 5.7 | 3.9 |
| 6 | 5.5 | 563.3 | 122.7 | 3.9 | 3.2 |
| 4 | 2 | 328.5 | 1.3 | 1.9 | 2.6 |
| 2 | 1.5 | 1 | 1.2 | 0.8 | 1.3 |
| 1 | 0.7 | 0.5 | 0.4 | 0.4 | 0.5 |

**Table 3. The average timing, i.e., the average number of weeks after the beginning of the simulated spreading, of all infection cases, for *p*= 1.0    (the cells in denser red color show the larger values in each column).**

$K = 1: N = 1000, p = 1.0$

| | | | $m_0$ | | |
|---|---|---|---|---|---|
| | 1 | 2 | 4 | 8 | 16 |
| 32 | 7.7 | 7.3 | 6.0 | 4.9 | 24.2 |
| 20 | 12.7 | 7.9 | 6.8 | 7.9 | 23.8 |
| 16 | 8.2 | 9.0 | 7.6 | 45.8 | 3.9 |
| 10 | 8.1 | 10.5 | 11.1 | 40.9 | 2.7 |
| 8 | 11.2 | 12.9 | 38.8 | 3.6 | 2.6 |
| 6 | 12.6 | 13.9 | 15.3 | 2.3 | 4.3 |
| 4 | 15.8 | 22.2 | 3.5 | 1.8 | 3.8 |
| 2 | 5.9 | 3.6 | 3.1 | 5.6 | 5.6 |
| 1 | 2.3 | 1.8 | 3.5 | 4.0 | 7.2 |

W (row label)

$K = 1: N = 10000, p = 1.0$

| | | | $m_0$ | | |
|---|---|---|---|---|---|
| | 1 | 2 | 4 | 8 | 16 |
| 32 | 14.2 | 10.5 | 9.0 | 8.2 | 44.2 |
| 20 | 16.0 | 13.7 | 10.3 | 12.0 | 43.6 |
| 16 | 14.8 | 14.0 | 10.3 | 39.0 | 3.7 |
| 10 | 14.0 | 16.1 | 17.4 | 42.0 | 2.4 |
| 8 | 17.5 | 17.0 | 40.4 | 3.4 | 3.3 |
| 6 | 17.9 | 22.8 | 21.1 | 2.4 | 2.7 |
| 4 | 17.6 | 16.1 | 2.7 | 2.5 | 2.9 |
| 2 | 7.1 | 3.1 | 3.4 | 2.5 | 2.0 |
| 1 | 2.0 | 2.8 | 1.5 | 3.6 | 2.5 |

$K = 0.1N/W: N = 1000, p = 1.0$

| | | | $m_0$ | | |
|---|---|---|---|---|---|
| | 1 | 2 | 4 | 8 | 16 |
| 32 | 6.6 | 9.3 | 6.1 | 4.1 | 8.8 |
| 20 | 10.1 | 10.3 | 6.5 | 4.6 | 23.8 |
| 16 | 11.2 | 10.9 | 6.1 | 9.7 | 3.5 |
| 10 | 6.1 | 12.0 | 7.7 | 24.9 | 2.6 |
| 8 | 6.9 | 11.1 | 12.3 | 3.3 | 3.1 |
| 6 | 6.5 | 13.8 | 11.0 | 5.2 | 2.9 |
| 4 | 5.8 | 25.9 | 3.2 | 2.4 | 4.3 |
| 2 | 5.9 | 3.4 | 2.4 | 3.9 | 6.4 |
| 1 | 3.3 | 2.6 | 1.4 | 5.4 | 8.6 |

$K = 0.1N/W: N = 10000, p = 1.0$

| | | | $m_0$ | | |
|---|---|---|---|---|---|
| | 1 | 2 | 4 | 8 | 16 |
| 32 | 9.7 | 10.3 | 7.4 | 6.9 | 10.0 |
| 20 | 14.7 | 13.5 | 8.3 | 8.1 | 26.4 |
| 16 | 8.8 | 12.9 | 11.0 | 12.1 | 2.4 |
| 10 | 7.7 | 17.1 | 11.7 | 9.7 | 3.3 |
| 8 | 10.1 | 19.7 | 15.8 | 3.5 | 3.1 |
| 6 | 4.8 | 20.6 | 15.2 | 1.9 | 2.3 |
| 4 | 4.5 | 34.2 | 1.8 | 1.8 | 3.0 |
| 2 | 6.2 | 2.8 | 3.3 | 2.5 | 2.4 |
| 1 | 2.8 | 2.1 | 2.4 | 2.0 | 2.1 |

**Table 4. The average timing, i.e., the average number of weeks after the beginning of the simulated spreading, of all infection cases, for *p*= 0.5    (the cells in denser red color show the larger values in each column) .**

$K = 1: N = 1000, p = 0.5$

| | | | $m_0$ | | |
|---|---|---|---|---|---|
| | 1 | 2 | 4 | 8 | 16 |
| 32 | 7.8 | 8.4 | 5.7 | 5.5 | 27.0 |
| 20 | 9.1 | 9.4 | 8.8 | 8.3 | 30.2 |
| 16 | 11.1 | 11.3 | 8.0 | 40.1 | 1.9 |
| 10 | 13.7 | 11.7 | 12.0 | 44.6 | 3.9 |
| 8 | 13.4 | 16.3 | 43.6 | 3.7 | 6.3 |
| 6 | 15.8 | 20.4 | 12.8 | 4.5 | 5.2 |
| 4 | 11.5 | 28.1 | 3.8 | 5.4 | 4.9 |
| 2 | 4.3 | 2.3 | 5.1 | 5.0 | 7.6 |
| 1 | 4.5 | 3.0 | 3.0 | 5.5 | 7.3 |

$K = 1: N = 10000, p = 0.5$

| | | | $m_0$ | | |
|---|---|---|---|---|---|
| | 1 | 2 | 4 | 8 | 16 |
| 32 | 14.4 | 12.8 | 10.4 | 10.2 | 42.8 |
| 20 | 19.8 | 14.5 | 11.8 | 14.1 | 43.8 |
| 16 | 19.2 | 15.3 | 10.5 | 43.9 | 2.7 |
| 10 | 16.1 | 18.3 | 17.9 | 37.9 | 5.2 |
| 8 | 18.4 | 22.9 | 39.8 | 3.9 | 4.9 |
| 6 | 22.4 | 28.6 | 14.4 | 3.1 | 3.2 |
| 4 | 21.4 | 23.3 | 3.4 | 3.6 | 3.0 |
| 2 | 5.2 | 3.3 | 3.6 | 4.2 | 4.5 |
| 1 | 3.6 | 3.3 | 4.6 | 5.4 | 3.3 |

$K = 0.1N/W: N = 1000, p = 0.5$

| | | | $m_0$ | | |
|---|---|---|---|---|---|
| | 1 | 2 | 4 | 8 | 16 |
| 32 | 8.2 | 9.5 | 6.6 | 5.0 | 9.9 |
| 20 | 11.7 | 12.4 | 8.3 | 5.4 | 31.8 |
| 16 | 3.3 | 11.9 | 9.7 | 10.5 | 4.7 |
| 10 | 11.8 | 13.6 | 9.3 | 31.3 | 4.4 |
| 8 | 9.3 | 13.2 | 12.9 | 3.8 | 4.6 |
| 6 | 6.1 | 21.7 | 15.3 | 4.1 | 3.5 |
| 4 | 7.3 | 31.0 | 3.7 | 3.9 | 4.6 |
| 2 | 8.3 | 4.7 | 3.8 | 7.6 | 7.6 |
| 1 | 3.2 | 3.1 | 2.5 | 5.5 | 4.4 |

$K = 0.1N/W: N = 10000, p = 0.5$

| | | | $m_0$ | | |
|---|---|---|---|---|---|
| | 1 | 2 | 4 | 8 | 16 |
| 32 | 11.9 | 13.3 | 8.8 | 7.7 | 11.1 |
| 20 | 17.8 | 14.8 | 10.3 | 8.8 | 26.2 |
| 16 | 7.5 | 17.0 | 10.9 | 12.3 | 2.1 |
| 10 | 7.8 | 19.6 | 12.8 | 16.1 | 2.2 |
| 8 | 10.7 | 22.3 | 19.3 | 2.7 | 2.4 |
| 6 | 8.9 | 25.4 | 15.6 | 2.4 | 2.8 |
| 4 | 5.2 | 43.6 | 3.2 | 2.4 | 2.2 |
| 2 | 5.9 | 2.5 | 2.2 | 4.4 | 2.0 |
| 1 | 2.1 | 3.0 | 2.4 | 2.9 | 2.5 |